\documentclass[prb,twocolumn,showpacs,amsmath,amssymb,preprintnumbers,superscriptaddress]{revtex4}
\usepackage{dcolumn}
\usepackage{bm}
\usepackage{graphicx}
\usepackage{color}

\begin{document}

\title{Low temperature ferromagnetism in perovskite SrIrO$_3$ films}

\author{Rachna Chaurasia}\affiliation{School of Physical Sciences, Jawaharlal Nehru University, New Delhi - 110067, India.}
\author{K. Asokan}\affiliation{Materials Science Division, Inter University Accelerator Centre, New Delhi- 110 067, India.}
\author{Kranti Kumar}\affiliation{UGC-DAE Consortium for Scientific Research, Indore - 452001, India.}
\author{A. K. Pramanik}\email{akpramanik@mail.jnu.ac.in}\affiliation{School of Physical Sciences, Jawaharlal Nehru University, New Delhi - 110067, India.}

\begin{abstract}
The 5$d$ based SrIrO$_3$ represents prototype example of nonmagnetic correlated metal which mainly originates from a combined effect of spin-orbit coupling, lattice dimensionality and crystal structure. Therefore, tuning of these parameters results in diverse physical properties in this material. Here, we study the structural, magnetic and electrical transport behavior in epitaxial SrIrO$_3$ film ($\sim$ 40 nm) grown on SrTiO$_3$ substrate. Opposed to bulk material, the SrIrO$_3$ film exhibits a ferromagnetic ordering at low temperature below $\sim$ 20 K. The electrical transport data indicate an insulating behavior where the nature of charge transport follows Mott's variable-range-hopping model. A positive magnetoresistance is recorded at 2 K which has correlation with magnetic moment. We further observe a nonlinear Hall effect at low temperature ($<$ 20 K) which arises due to an anomalous component of Hall effect. An anisotropic behavior of both magnetoresistance and Hall effect has been evidenced at low temperature which coupled with anomalous Hall effect indicate the development of ferromagnetic ordering. We believe that an enhanced (local) structural distortion caused by lattice strain at low temperatures induces ferromagnetic ordering, thus showing structural instability plays vital role to tune the physical properties in SrIrO$_3$.       
\end{abstract}


\maketitle
\section{Introduction}
In recent times, lots of scientific interest have been placed on Ir-based transition metal oxides.\cite{kim1,kim2,krempa,cao1,blanchard1,longo,moon,rau} Due to its 5$d$ character, iridates exhibit strong spin-orbit coupling (SOC) and relatively weak electron correlation ($U$) effect. The delicate balance among competing energies such as, SOC, $U$ and crystal field effect (CFE) gives exotic electronic and magnetic properties in these materials where many of the phases are topologically relevant. In presence of strong SOC, the $t_{2g}$ $d$-orbitals split into low lying $J_{eff}$ = 3/2 and top lying $J_{eff}$ = 1/2 states. In case of Ir$^{4+}$ with 5$d^5$ electronic configuration, the $J_{eff}$ = 3/2 state is fully filled while $J_{eff}$ = 1/2 remains half filled. The later $J_{eff}$ state being narrow, even a small $U$ opens up a gap, thus giving a realization of $J_{eff}$ = 1/2 Mott-like insulating state.\cite{kim1,kim2}

The in-built structural dimensionality has, however, a dominant role on the magnetic and electronic properties of iridium oxide materials. For instance, in Ruddlesden-Popper (RP) series Sr$_{n+1}$Ir$_n$O$_{3n+1}$ (which can be considered as SrO.(SrIrO$_3$)$_n$ where $n$ layers of perovskite SrIrO$_3$ is separated by magnetically and electronically inactive SrO layer), a transition from paramagnetic (PM) and metallic state in perovskite SrIrO$_3$ ($n$ =  $\infty$) to magnetic and insulating state has been observed in layered Sr$_2$IrO$_4$ ($n$ = 1) and Sr$_3$Ir$_2$O$_7$ ($n$ = 2).\cite{blanchard1,pallecchi,cao2,bhatti} Among the iridium oxides, the perovskite SrIrO$_3$ draws particular interest as it is shown to lie on the verge of ferromagnetic (FM) instability and metal-insulator transition (MIT).\cite{zeb} In addition, the low temperature electronic state in SrIrO$_3$ has been characterized with a non-Fermi-liquid behavior.\cite{cao3} Therefore, tuning of parameters, such as SOC, $U$, structural distortion, etc. by means of doping or lattice strain likely to modify its ground state electronic and magnetic properties significantly. In fact, a recent band structure calculation shows a line node near the Fermi surface which is inherent to crystal structure of this material, and predicts SrIrO$_3$ can host topological phases upon tuning the SOC strength with suitable doping.\cite{carter} Further, an emergence of antiferromagnetic (AFM) and insulating state with the substitution of a nonmagnetic and isovalent element in SrIr$_{1-x}$Sn$_x$O$_3$ highlights the role of structural distortion driven tuning of its physical properties.\cite{cui}   

In present work, we have studied the magnetic and electrical transport behavior in epitaxial SrIrO$_3$ film ($\sim$ 40 nm) grown on SrTiO$_3$ substrate. Bulk SrIrO$_3$ adopts a hexagonal structure grown in an ambient condition but a perovskite orthorhombic crystal structure can be realized if synthesized under high pressure.\cite{longo,zhao} However, an epitaxial film grown on suitable substrate has an advantage that can host an orthorhombic SrIrO$_3$. Moreover, substrate strain can be used as a tuning parameter to get modified magnetic and electronic behavior. There have been several studies for SrIrO$_3$ film with different substrates. An angle-resolved photoemission spectroscopy (ARPES) study on perovskite SrIrO$_3$ film reveals a narrow semi-metallic band across Fermi level which mainly originates due to combined effect of SOC, dimensionality and IrO$_6$ octahedral rotations.\cite{nie} The effects of substrate strain, film thickness and substrate temperature on metal-insulator transition in perovskite SrIrO$_3$ film have been studied.\cite{zhang1,biswas1,biswas2,bhat} For SrIrO$_3$ multilayers, many exotic phenomena such as, topological Hall effect, anomalous Hall effect, tuning of magnetic anisotropy, exchange bias, etc. have also been studied in heterostructures consisting of SrIrO$_3$ and FM oxides.\cite{matsuno,nichols,chaurasia,yi} While most of the studies have focused on exotic transport and electronic properties, a less attention has been paid to the evolution of magnetic behavior in SrIrO$_3$ films. A dimensionality controlled magnetic and electronic properties have recently been studied in artificial [(SrIrO$_3$)$_m$,(SrTiO$_3$)$_n$] ($m$, $n$ = 1, 2, 3,.....) superlatices.\cite{matsuno1,hao} These studies show a low temperature magnetic ordering in SrIrO$_3$ layer with the transition temperature around 140 K for $m$, $n$ = 1 which systematically decreases with increasing $m$, and further show a close relation of resistivity with the magnetic transition. A recent theoretical calculation shows that the magnetic state in bulk perovskite SrIrO$_3$ is significantly modified with the tuning of SOC.\cite{zeb} Therefore, the lattice strain or the distortion of local IrO$_6$ octahedra in films likely to play vital role in determining its magnetic behavior. The prominent example of strain induced ferromagnetism is CaRuO$_3$ film where its bulk component has similar orthorhombic structure showing similar non-magnetic and non-Fermi-liquid behavior.\cite{tripathi}

The present SrIrO$_3$ film is found to be epitaxial with good crystal quality. While the magnetic data indicate a development of (weak) FM ordering at low temperature ($<$ $\sim$ 20 K), the film remains insulating all over the temperature. A positive, though small magnetoresistance (MR) is observed at 2 K which has correlation with its magnetic behavior. The nonlinear Hall effect coupled with anisotropic-MR and -Hall effect further supports an evolution of FM state at low temperature. An increasing strain or local structural distortion at low temperature is believed to induce the FM behavior.      
         
\section{Experimental Methods}
Epitaxial thin film of SrIrO$_3$ with thickness $\sim$ 40 nm has been grown on (100) oriented SrTiO$_3$ single-crystal substrate using pulsed laser deposition (PLD) technique equipped with KrF ($\lambda$ = 248 nm) laser. A phase-pure stoichiometric SrIrO$_3$ polycrystalline pellet is used as a target for film deposition. Before deposition, the substrate has been properly cleaned in ultrasonic cleaner alternatively with acetone and isopropyl alcohol for about 10 min. The laser frequency and energy are used as 5 Hz and 1.5 J/cm$^2$, respectively. The deposition has been done with parameters such as, substrate to target distance 5 cm, substrate temperature 750 $^{\circ}$C and oxygen pressure during deposition 0.1 mbar. In order to maintain the oxygen stoichiometry, the deposited films have been post annealed at same deposition temperature 750 $^{\circ}$C for about 15 min at partial oxygen pressure around 500 mbar. The thickness of the films has been estimated using calibrated laser shot counts\cite{kharkwal} which closely matches with the thickness checked with FESEM. The structural characterization of the films are performed with x-ray diffraction (XRD) where the data have been collected using a Panlytical diffractometer equipped with Cu-$K_{\alpha}$ source. X-ray absorption spectroscopy (XAS) data have been collected from ` National synchrotron radiation research center’, Taiwan for O-K edge and Ir-$L_3$ edge in total electron yield and fluorescence modes, respectively by following standard procedure. Before XAS measurements, the x-ray photon energy has been calibrated using a metallic gold foil for $L_3$ edge absorption which is the standard procedure followed in beamline. Further, XAS data are collected for various iridate samples having different Ir charge states in same beamtime and same beamline which provides a better comparison. The electronic transport and its angle dependent measurements are done using a four-probe method in an insert attached with 9 Tesla magnet (Nanomagnetic). Both temperature and magnetic field dependent magnetic properties of the film are measured with a superconducting quantum interference device (Quantum Design). The magnetic contribution due to film has been extracted after subtracting the related moment of substrate.    

\begin{figure}
\centering
		\includegraphics[width=8cm]{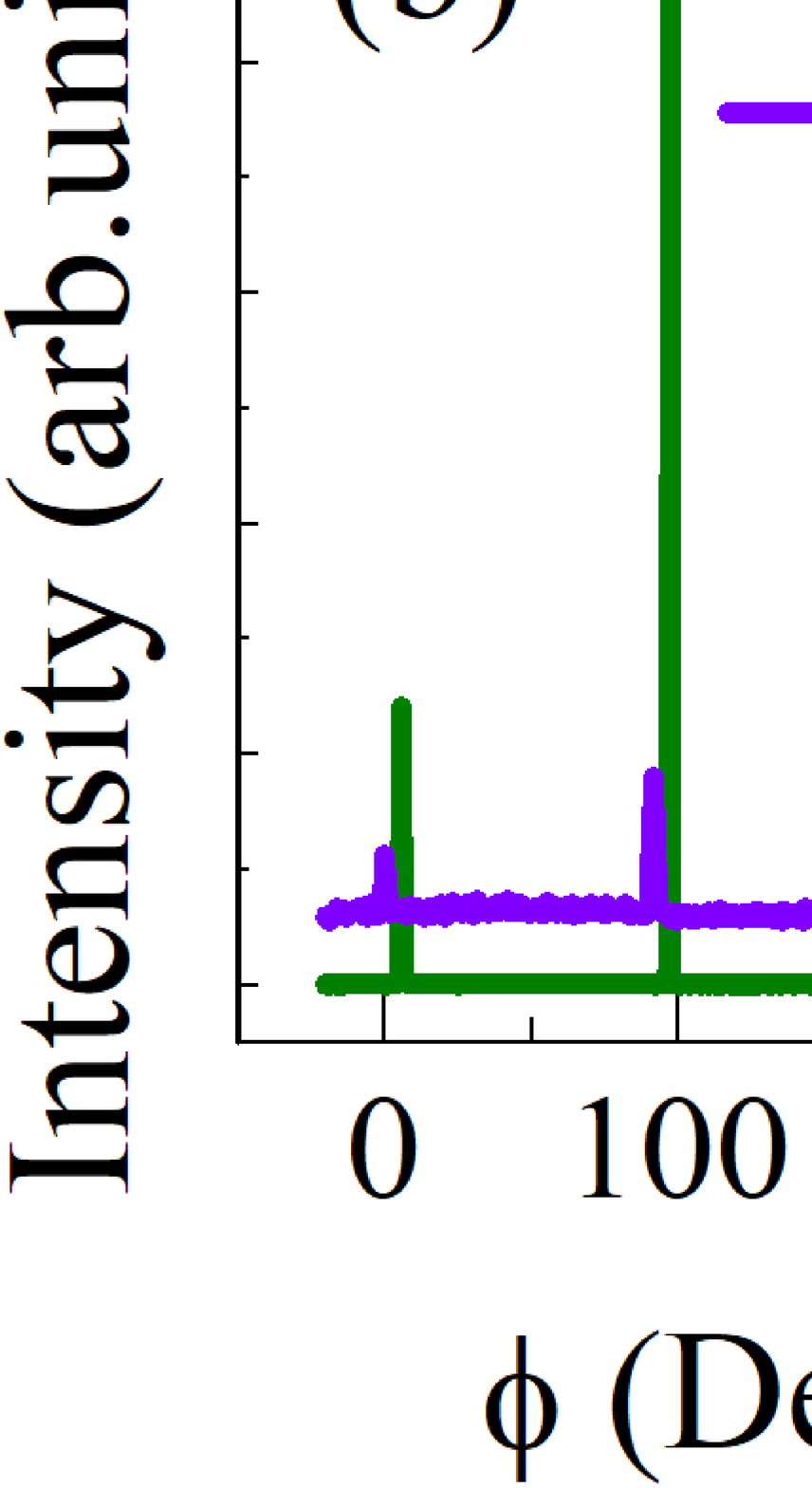}
\caption{(color online) (a) the XRD pattern of SrIrO$_3$ film grown on SrTiO$_3$ (100) substrate are shown in semi-log scale. Left inset shows the expanded (100) Bragg peak with thickness fringes while the right inset presents an atomic force microscope image of the film showing surface topography. (b) shows the $\phi$-scan at (220) reflection. (c) shows the $\omega$-scan (rocking curve) of SrIrO$_3$ film around (200) reflection.}
	\label{fig:Fig1}
\end{figure}

\section{Results and Discussion}
\subsection{Structural Analysis with x-ray diffraction}
Fig. 1a shows $\theta$-2$\theta$ x-ray diffraction (XRD) plot of SrIrO$_3$ film deposited on SrTiO$_3$ (100) substrate. Bulk SrIrO$_3$ is realized from Ruddlesden-Popper series Sr$_{n+1}$Ir$_{n}$O$_{3n+1}$ with $n$ = $\infty$ where an infinite layers of perovskite SrIrO$_3$ are stacked together forming a 3-dimensional structural network. Bulk SrIrO$_3$ generally adopts two different crystal structures based on synthesis protocol. At ambient pressure, SrIrO$_3$ crystallizes in 6$H$-hexagonal (monoclinic) structure while using high pressure synthesis method this material stabilizes in perovskite orthorhombic ($Pbnm$) structure. As a target material for deposition of present film, single phase SrIrO$_3$ is used which has been synthesized at ambient pressure having monoclinic-\textit{C2/c} structure with lattice parameters $a$ = 5.5982 {\AA}, $b$ = 9.6293 {\AA}, $c$ = 14.1949 {\AA} and $\beta$ = 93.228$^o$. The epitaxial growth of films has, however, advantage that a meta-stable orthorhombic phase can be stabilized with perovskite substrate. The used substrate SrTiO$_3$ has cubic structure with lattice parameter $a_{sub}$ = 3.90 \AA. The pseudo-cubic ($pc$) lattice parameter $a_{pc}$ ($\approx$ 0.5$\sqrt{a^2+b^2}$) of target SrIrO$_3$ has been calculated from its bulk orthorhombic lattice parameters ($a$ = 5.56 {\AA}, $b$ = 5.59 {\AA} and $c$ = 7.88 {\AA}) giving its value 3.942 {\AA} which corresponds to $\sim$ +1\% compressive lattice strain for the films deposited on SrTiO$_3$. Given that $a_{pc}$ and $a_{sub}$ has slight mismatch while the lattice parameter $c$ of orthorhombic SrIrO$_3$ matches closely with 2$a_{sub}$, the SrIrO$_3$ films on SrTiO$_3$ (100) substrate are likely to grow along (110) direction rather than having (100) orientation.

\begin{figure}
\centering
		\includegraphics[width=8cm]{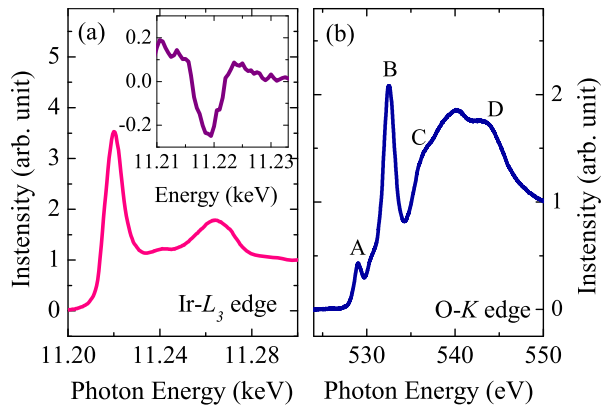}
\caption{(color online) (a) XAS spectra at Ir-$L_3$ edge and (b) XAS spectra at O-$K$ edge of the SrIrO$_3$ film are shown at room temperature. Inset of (a) shows the double derivative of XAS spectra of Ir-$L_3$ edge. A, B, C and D in (b) mark the energy position for for the peak/hump.}
	\label{fig:Fig2}
\end{figure}
 
As shown in main panel of Fig. 1a, the XRD pattern of deposited film exhibit only crystalline peaks without trace of any impurity or additional phase(s). A satellite peak near the substrate peak signifies the epitaxial growth of the SrIrO$_3$ film. A magnified view of low-angle (100) reflection has been shown in left inset of Fig. 1a with thickness fringes. While the peaks due to substrate and film are not superimposed but they are very close which implies the film is under strain. The Bragg peaks related to SrTiO$_3$ and SrIrO$_3$ are observed at 2$\theta$ = 22.74$^{\circ}$ and $\theta$ = 22.44$^{\circ}$ which gives the lattice parameter 3.90 {\AA} and 3.96 {\AA}, respectively. The calculated lattice parameter of SrIrO$_3$ film is very close to its bulk $a_{pc}$ (3.942 {\AA}). The SrTiO$_3$ substrate in reality gives a compressive strain $\sim$ +1.5\% to SrIrO$_3$ film.

Further, an estimation regarding the thickness ($D$) of deposited film has been done using thickness fringes in (100) reflection with the following formula,

\begin{eqnarray}
 D = \frac{(m-n)\lambda}{2(Sin \theta_m - Sin \theta_n)}
\end{eqnarray}

where $\theta_m$ and $\theta_n$ are the positions of $m$-th and $n$-th order peaks and $\lambda$ is the wavelength of x-ray used. Following Eq. 1, we have calculated the thickness of SrIrO$_3$ film $\sim$ 39.7 nm which is close to expected value ($\sim$ 40 nm) from growth rate estimation.

An AFM image of film surface is shown in right inset of Fig. 1a. The AFM image shows no voids but an average surface roughness $\sim$ 1 nm. To further understand the structure of deposited film, a $\phi$-scan has been taken at (220) reflection for both film as well as SrTiO$_3$ substrate, as shown in Fig. 1b. The SrTiO$_3$ has four-fold symmetric cubic structure with in-plane and out-of-plane pseudo-cubic lattice parameter 3.90 {\AA}. As seen in figure, the SrTiO$_3$ and SrIrO$_3$ $\phi$-scan peaks are nearly equidistant with $\sim$ 90$^{\circ}$ apart, and the peaks are close to each other having a very small difference with $\Delta \phi$ $\sim$ 5.5$^{\circ}$. This suggests a four-fold structural symmetry of SrIrO$_3$ film and the film has single domain with cube on cube growth. Moreover, an unsplit nature of peaks in $\phi$-scan data (Fig. 1b) is in favor orthorhombic structure rather than monoclinic one.\cite{bhat} The $\theta$-2$\theta$ XRD pattern and the $\phi$-scan underlines the fact that the present SrIrO$_3$ film has taken an orthorhombic structure on SrTiO$_3$ substrate. Our result is in line with previous reports which has shown that the film thinner than 40 nm would take orthorhombic structure while a thicker film is susceptible to mixed phase of monoclinic and orthorhombic structures.\cite{zhang2,jang} We have further characterized the crystalline quality of deposited SrIrO$_3$ film with a $\omega$-scan (rocking curve) in XRD measurements. Fig. 1c shows the $\omega$-scan of present SrIrO$_3$ film, obtained around (200) reflection. The full width half maxima (FWHM) of $\omega$-scan has been calculated to be around 0.5$^{\circ}$ which compares well with other report of similarly thick films.\cite{biswas2,bhat} The $\omega$-scan usually signifies about the perfection in lattice planes and mosaic spread, therefore reasonably small FWHM suggests the present film has parallel planes and less mosaic spread. All these results conclusively show that the present SrIrO$_3$ film is high quality epitaxial film with an orthorhombic structure, obtained on SrTiO$_3$ substrate.

\subsection{X-ray absorption spectroscopy}
To understand the Ir oxidation state as well as Ir-O hybridization in present film, x-ray absorption spectroscopy (XAS) measurements have been done at room temperature. Fig. 2a shows normalized $L_3$ (2$p_{3/2}$ $\rightarrow$ 5$d$) absorption edge spectra for the SrIrO$_3$ film which involves transition to both 5$d_{5/2}$ and 5$d_{3/2}$ states. It is evident in Fig. 2a that $L_3$ edge occurs at 11219.5 eV. Usually, in XAS spectra the position of absorption edge largely depends on the ionic state of transition metal because the transition metal - oxygen bond length mostly determines the energy shifting of absorption edge. With an increase in ionic state the bond-length becomes shorted, therefore the energy edge occurs at higher energy. However, a small difference in edge position has been observed even for same ionic state with different lattice structure. In case of thin films where the lattice strain has significant role on the bond length, the position of absorption edge may vary with the bulk material. In case of iridium, the Ir-$L_3$ absorption edge has usually been seen to occur at 11218.0, 11219.6, 11220.0, 11222.0 and 11222.5 eV for Ir, Ir$^{3+}$, Ir$^{4+}$, Ir$^{5+}$ and Ir$^{6+}$ charge state, respectively.\cite{laguna,zhou,liu,harish,clancy,kharkwal1} Given that the absorption edge is very sensitive to local environment of anions, a small difference in energy position may occur from system to system. The $L_3$ edge at 11219.95(4) eV in present film suggests iridium is in Ir$^{4+}$ state. However, a little lower value of edge position compared to 11220 eV may be due to strain effect in film giving a modified Ir-O bond length or due to presence of small fraction of Ir$^{3+}$ related to oxygen vacancy during film deposition. While there is difference in peak position even for same element in different materials which is mainly due to different chemical environment and composition,\cite{laguna,clancy,harish} the fact is that absorption edge increases with increasing charge state. In this sense, the $L_3$ XAS data imply a Ir$^{4+}$ charge state in present film. To check whether there is any mixing of Ir oxidation states, we have plotted double derivative in inset of Fig. 2a. A distinct shoulder in double derivative of $L_3$ spectra is generally considered to be an indicative of mixed charge state,\cite{laguna} but no clearly visible shoulder in double derivative of our $L_3$ spectra is observed, as shown in inset of Fig. 2a. Thus, based on this results we conclude that iridium mostly is in Ir$^{4+}$ oxidation state.

We have additionally measured XAS spectra at O-$K$ edge on present film to understand about hybridization between Ir-5$d$ and O-2$p$ states. Following crystal field chemistry, the Ir-$d$ orbitals in environment of IrO$_6$ octahedra are split into low-lying $t_{2g}$ ($d_{xy}$, $d_{yz}$ and $d_{zx}$) and high-lying $e_g$ ($d_{x^2-y^2}$ and $d_{z^2}$) states. In octahedral environment, the oxygen orbitals ($p_x$, $p_y$ and $p_z$) of six ligands (four basal and two apical) hybridize with transition metal $d$ orbitals. Among $t_{2g}$ orbitals, $d_{xy}$ hybridizes only with basal $p_x$/$p_y$ while $d_{yz}$ and $d_{zx}$ hybridize with both basal $p_z$ and apical $p_x$/$p_y$ orbitals. In case of $e_g$ orbitals, $d_{x^2-y^2}$ and $d_{z^2}$ engage in interaction with $p_x$/$p_y$ and $p_z$, respectively. The normalized O-$K$ edge spectra for present film in Fig. 2b shows two distinct peaks (marked by A and B in plot) in lower energy regime at binding energy ($E_b$) around 529 and 532.5 eV, respectively while a broad hump is observed between 534 and 550 eV. The onset energy of present O-$K$ spectral edge agree with other report for SrIrO$_3$ film.\cite{liu} The peak at 529 eV (A marking) is due to hybridization between Ir-$d_{xz}$/$d_{yz}$ and apical O-$p_x$/$p_y$ while that at 532.5 eV (B marking) arises due to interaction between Ir-$t_{2g}$ and and basal O-$p$ orbitals. On the other hand, the hybridization between Ir-$e_g$ and O-$p$ orbitals causes broad hump in higher energy side. It can be further noticed in Fig. 2 that the broad hump in high energy side exhibits two prominent shoulders (marked by C and D in figure). While the above hybridization picture has been discussed considering an isolated model of $e_g$ and $t_{2g}$ orbitals, a recent theoretical study has, however, shown that in 5d based oxides, the crystal field effect, SOC and $U$ has prominent role on mixing of $e_g$ and $t_{2g}$ orbitals, hence influences the absorption edge spectra.\cite{stam} Further, note that we observe a slight difference in peak positions compared to bulk materials, which is probably due to substrate induced strain in films which alters the extent of hybridization through local structural modification. Such modification in strength of hybridization vis-$\grave{a}$-vis peak position in XAS O-$K$ edge spectra has been observed with chemical pressure in (Y$_{1-x}$Pr$_x$)$_2$Ir$_2$O$_7$.\cite{harish}

\begin{figure}
\centering
		\includegraphics[width=8.5cm]{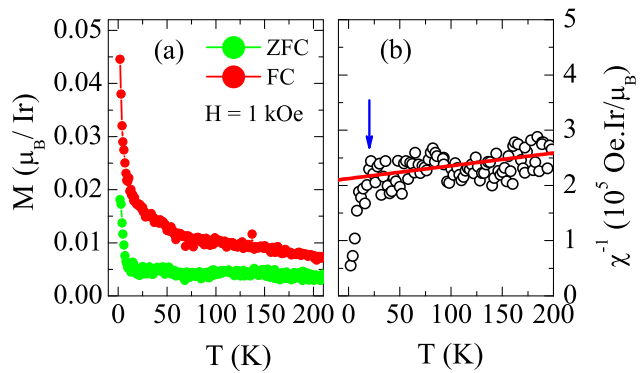}
\caption{(color online) (a) Temperature dependent magnetization data measured in 1 kOe applied field following ZFC and FC protocol are shown for SrIrO$_3$ film. (b) shows the inverse magnetic susceptibility $\chi^{-1}$ (= $(M/H)^{-1}$) as a function of temperature, deduced from ZFC magnetization data. The red straight line is due to fitting of straight line following Curie-Weiss law (discussed in text) while the arrow indicates the deviation from linearity.}
	\label{fig:Fig3}
\end{figure}

\subsection{Magnetization Study}
Fig. 3a shows the temperature dependent magnetization ($M$) data for SrIrO$_3$ film, collected in applied field of 1 kOe following zero field cooling (ZFC) and field cooling (FC) protocol. The moment of the film has been extracted after subtracting the substrate contribution. On cooling, while ZFC and FC branches of magnetization exhibit a finite difference starting from high temperature but the moment in both measurements increases monotonically till about 20 K. Below 20 K, both $M_{FC}$ and $M_{ZFC}$ show a sharp increase till lowest measurement temperature of 2 K. This sharp increase in moment below $\sim$ 20 K implies a development of weak ferromagnetism in SrIrO$_3$ film at low temperature. However, the moment of the film is quite low which is in line with the fact that iridates are generally low moment systems. Note, that both the low temperature FM state as well as the obtained moment are in agreement with artificial [(SrIrO$_3$)$_m$,(SrTiO$_3$)$_n$] superlatices.\cite{matsuno1,hao} Further, the observed magnetic behavior of SrIrO$_3$ film is consistent with our recent report of exchange bias behavior in La$_{0.67}$Sr$_{0.33}$MnO$_3$/SrIrO$_3$ multilayers at low temperature below $\sim$ 20 K.\cite{chaurasia} Note, that similar sharp rise in susceptibility has also been observed in bulk SrIrO$_3$ below $\sim$ 15 K which has been shown due to proximity to FM instability.\cite{cao3} At low temperatures, there is significant exchange enhancement, though a FM ordering is not developed which probably requires a triggering. In fact, similar ferromagnetic instability has been predicted in theoretical calculations for bulk SrIrO$_3$,\cite{zeb} hence any suitable perturbation likely to trigger the magnetism in this material. Along with chemical doping,\cite{cui,qasim} the lattice strain arising from underlying substrate acts as driving force to induce magnetic state in films. Further, the inverse magnetic susceptibility $\chi^{-1}$, calculated as ($M_{ZFC}/H)^{-1}$, are shown in inset of Fig. 3b as a function of temperature. As evident in figure, above 20 K the $\chi^{-1}(T)$ shows a linear increase where the behavior is similar to Curie-Weiss (CW) behavior, $\chi$ = $C/(T-\theta_P$) where $C$ and $\theta_P$ are the Curie constant and Curie temperature, respectively. However, the fitting parameters such as, $C$ and $\theta_P$ can not be determined precisely. Nonetheless, the deviation from linear behavior of $\chi^{-1}(T)$ below around 20 K suggests an onset of FM at low temperature.

\begin{figure}
\centering
		\includegraphics[width=8cm]{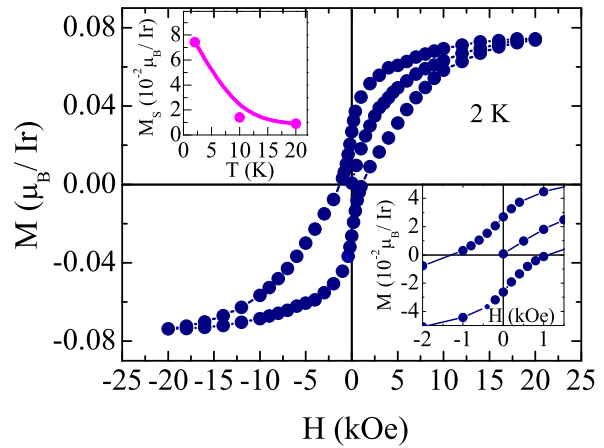}
\caption{(color online) Magnetic field dependent magnetization loop is shown for SrIrO$_3$ thin film at 2 K. The upper inset shows the variation of saturated magnetization $M_s$ with temperature. The lower inset depicts an expanded view of $M(H)$ close to origin showing an asymmetry in magnetic hysteresis loop.}
	\label{fig:Fig4}
\end{figure}
 
To understand the low temperature magnetic state further, we have recorded magnetic hysteresis loop $M(H)$ at three different temperature 2, 10 and 20 K with field range of $\pm$ 20 kOe. The magnetic contribution due to SrTiO$_3$ substrate has been subtracted from original $M(H)$ data at each temperature. We have adopted the protocol where a slope in $M(H)$ data has been taken at high field regime.  Assuming this slope represents the susceptibility of substrate, moment of the substrate has been calculated which has been used for subtraction to obtain the moment of film. Fig. 4 shows the representative corrected $M(H)$ plot for present SrIrO$_3$ film at 2 K. Unlike bulk SrIrO$_3$, the $M(H)$ plot shows an open hysteresis where we find left and right coercive field $H_c^L$ = 1293 Oe, $H_c^R$ = 1109 Oe, and upper and lower remnant magnetization $M_r^U$ = 2.69 $\times$ 10$^{-2}$ $\mu_B$/Ir, $M_r^L$ = 2.63 $\times$ 10$^{-2}$ $\mu_B$/Ir, respectively. This asymmetry in $M(H)$ plot ($H_c^L$ $\neq$ $H_c^R$), which is clearly shown in lower inset of Fig. 4, generally arises due to an exchange bias (EB) effect. For this present film, we calculate an exchange bias field $H_{EB}$ (= ($|H_c^L|$ + $|H_c^R|$)/2) around 92 Oe at 2 K which reduces to about 81 Oe at 10 K. Usually, EB effect is realized when a system with FM/AFM interface is cooled in presence of magnetic field from high temperature and $M(H)$ is measured at low temperature.\cite{nogues} The applied field biases the exchange interaction at interface which results in a shifting of $M(H)$ plot or EB phenomena.\cite{chaurasia,kharkwal,ke} However, the interfaces in films or multilayers experience various other factors, such as structural distortion, electronic reconstruction, strain, etc. which modify the magnetic characters at interface accordingly.\cite{tokura,zubko} These in turn induces EB effect which has shown many many interesting properties. For instance, EB effect has been observed even in zero-field-cooled $M(H)$ plot\cite{chaurasia,kharkwal} and in single-layer magnetic films,\cite{sow} which are rather unusual. The observed EB effect in present single layer SrIrO$_3$, deposited on diamagnetic SrTiO$_3$, is likely due to strain effect. We speculate that due to an enhanced strain at interface, some SrIrO$_3$ layers adjacent to interface may achieve an antiferromagnetic ordering which in contact with rest ferromagnetic SrIrO$_3$ layers results in an exchange bias effect.

Fig. 4 further shows a saturation in moment within applied field of 20 kOe. The temperature variation of saturation moment $M_s$ is shown in upper inset of Fig. 4. We have obtained a very low $M_s$ which, however, agrees well with other studies of [(SrIrO$_3$)$_m$,(SrTiO$_3$)$_n$] superlatices.\cite{matsuno1,hao} While iridates generally exhibit a low moment, the obtained $M_s$ is much lower than the expected saturation moment 0.33 $\mu_B$/Ir, which can be calculated using $M_s$ = $g_J J_{eff} \mu_B$ with $g_J$ = 2/3 and $J_{eff}$ = 1/2 for strong SOC dominated systems. For instance, at 2 K we obtain saturation moment around 7.5 $\times$ 10$^{-2}$ $\mu_B$/Ir (upper inset in Fig. 4) which is roughly one order lower than the expected value following the $J_{eff}$ model. A detail investigation involving the microscopic tools is required to understand the nature of low temperature magnetism in present film. However, at this stage we can speculate that the nature of this low temperature magnetic state may be either of FM type without the $J_{eff}$ state or a canted-AFM type with a $J_{eff}$ state. It can be noted that a weak FM has been observed in layered iridate Sr$_2$IrO$_4$ due to canted-AFM ordering which is driven by Dzyaloshinskii-Moriya (DM) type antisymmetric interaction in this SOC dominated system.\cite{kim1,kim2} While rest of the results and analysis suggest a FM ordering at low temperature in present film (discussed later), with the progress of time the $J_{eff}$ model has been shown to deviate considerably in non-ideal (or distorted) octahedral arrangement in both theoretical and experimental studies. A recent theoretical study has even discussed the effect of CFE, SOC and $U$ on the mixing of $e_g$ and $t_{2g}$ states in 5$d$ oxides.\cite{stam} Very recently, an evolution of both spin and orbital moment has been shown with the lattice distortion in 3$d$-5$d$ double perovskite (Sr$_{1-x}$Ca$_x$)$_2$FeIrO$_6$.\cite{kharkwal3} Therefore, the magnetism in iridates continues to be an interesting subject of research.

Nonetheless, the combination of EB effect, open hysteresis loop in $M(H)$ plot and the magnetic saturation in $M(H)$ above 10 kOe imply the ferromagnetic nature of present SrIrO$_3$ film at low temperature. The thermal demagnetization of the saturation moment, as shown in upper inset of Fig. 4, further indicates low temperature ferromagnetic in SrIrO$_3$ film. Opposed to paramagnetic bulk SrIrO$_3$, this development of ferromagnetism in its films is quite noteworthy. We believe that epitaxial lattice strain in films plays important role in stabilizing FM state which has been similarly observed in other nonmagnetic perovskite material CaRuO$_3$.\cite{tripathi}

\begin{figure}
\centering
		\includegraphics[width=8cm]{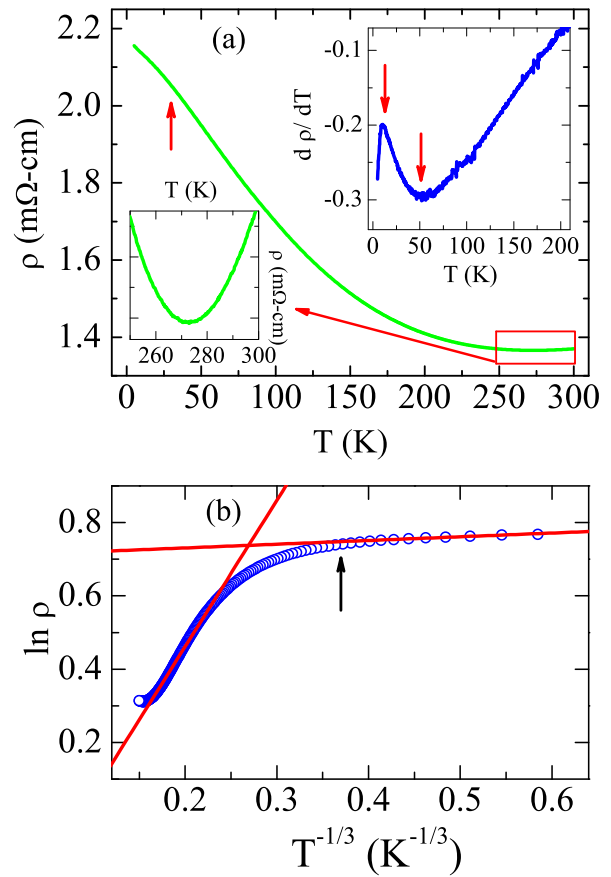}
\caption{(color online) (a) The electrical resistivity as a function of temperature are shown for SrIrO$_3$ film deposited on SrTiO$_3$ substrate. The up arrow marks an anomaly in resistivity around its magnetic transition. Upper inset shows temperature derivative of resistivity data showing the anomaly at low temperature as indicated by down arrows. Lower inset shows a magnified view of metal to insulator transition around 272 K. (b) shows the fitting of data with Mott variable-range-hopping model (Eq. 2) in two different temperature ranges. The solid lines are due to straight line fittings with Eq. 2. The arrow indicates the deviation from linearity in fitting in low temperature around 20 K.}
	\label{fig:Fig5}
\end{figure}

\subsection{Temperature dependent electronic transport measurements}
To understand the electronic transport behavior in present SrIrO$_3$ film, we have measured temperature dependent resistivity $\rho(T)$, as shown in Fig. 5. The resistivity value at room temperature is found to be $\sim$ 1.5 m$\Omega$-cm which is in good agreement with earlier reports (1 -2 m$\Omega$-cm).\cite{zhang1} With decreasing temperature, the $\rho(T)$ increases monotonically indicating a semi-metallic or insulating behavior. The electronic transport behavior in SrIrO$_3$ films are shown to be extremely sensitive to the deposition temperature, film thickness and lattice strain.\cite{zhang1,biswas1,biswas2,bhat} Previous studies with varying substrate temperature (500 - 800 $^o$C) have shown that with increasing substrate temperature, the film resistivity increases leading to an insulating behavior which is mainly due to an inhomogeneous Ir distribution.\cite{biswas1} Further, epitaxial strain realized either from reducing film thickness or using lattice mismatch substrates, has prominent role to increase the resistivity as well as to induce the insulating state. Given that our substrate deposition temperature is 750 $^o$C and the substrate SrTiO$_3$ has some lattice mismatch with SrIrO$_3$, this semi-metallic or insulating behavior is a likely behavior. However, an overall low resistivity and a reasonably low $\rho$(5 K)/$\rho$(300 K) ratio ($\sim$ 1.4) suggest a metal-like transport as for its bulk counterpart.

Moreover, a close inspection reveals a dip in $\rho(T)$ data around 272 K which has been shown in lower inset of Fig. 5. This dip in $\rho(T)$ appears to be a metal-insulator transition (MIT) which has been similarly observed at different temperatures in SrIrO$_3$ film with different substrates as well as with different film thickness.\cite{biswas2,zhang1,bhat} Here, it can be noted that below this dip, the $\rho(T)$ does not follow any logarithmic temperature dependance which is typical to the Kondo phenomenon. On cooling, a change in slope in $\rho(T)$ is further evident at low temperature around 30 K which is marked by an up arrow in Fig. 5. This change of slope in $\rho(T)$ occurs around its magnetic transition which is prominently observed in its temperature derivative, $d\rho$/$dT$, where a maximum slope change is observed around 50 and 10 K (marked by down arrow), as shown in upper inset of Fig. 5. The dip and peak in $d\rho$/$dT$ around 50 and 10 K, respectively suggest the resistivity changes with relatively faster rate between these temperatures. However, there is a disagreement between the dip temperature in $d\rho$/$dT$ around 50 K and the magnetic ordering temperature in Fig. 3 which may be due to an early onset of magnetic ordering. Nonetheless, the anomaly in $\rho(T)$ at low temperature is connected with the magnetic ordering which has been similarly observed in [(SrIrO$_3$)$_m$, (SrTiO$_3$)$_n$] ($m$, $n$ = 1, 2, 3,.....) superlatices.\cite{matsuno1,hao}

The $\rho(T)$ data can be fitted using Mott's variable-range-hopping (VRH) model,\cite{mott}

\begin{eqnarray}
\rho = \rho_0\exp[(T_0/T)^\alpha]
\end{eqnarray}

\begin{eqnarray}
  T_0 = \frac{21.2}{k_B N(\epsilon_F) \xi^3}
\end{eqnarray}

\begin{figure}
\centering
		\includegraphics[width=8cm]{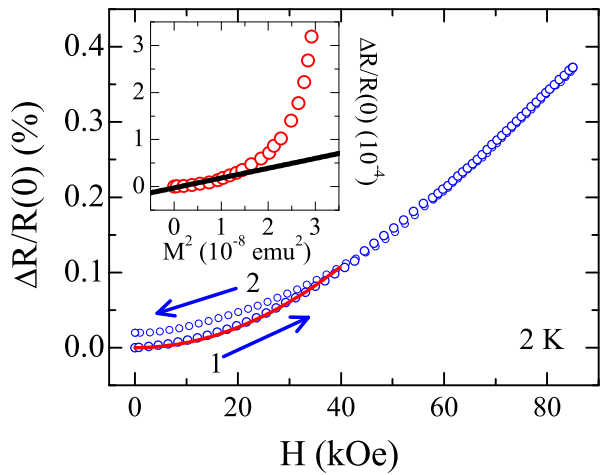}
\caption{(color online) Magnetoresitance (defined in text) as a function of magnetic field are shown for SrIrO$_3$ film at 2 K with only positive field direction. The arrows indicate the direction of field application. The solid line is due to fitting of MR data with quadratic field dependance. Inset shows the quadratic dependance of magnetoresitance on magnetization.}
	\label{fig:Fig6}
\end{figure}

where $\alpha$ is related to the dimensionality ($d$) of system with $\alpha$ = 1/($d$+1), $T_0$ is the characteristic temperature related to the leakage rate of localized states at Fermi level, $k_B$ is the Boltzmann constant, $N(\epsilon_F)$ is the density of states (DOS) at Fermi level and $\xi$ is the localization length.\cite{wu} As shown in Fig. 5b, Eq. 2 can be fitted with $\rho(T)$ in two distinct temperature regimes (5 - 15 K and 67 - 205 K) with $d$ = 2, giving T$_0$ values 3.7 $\times$ 10$^{-4}$ K and 22.56 K, respectively. The $T_0$ exhibits distinctly low value in low temperature magnetic state. The $T_0$ is inversely proportional to both $N(\epsilon_F)$ and $\xi$ (Eq. 3). Given that this is an insulating system, so an increase of $N(\epsilon_F)$ at low temperature is very unlikely. Therefore, this change in $T_0$ can be explained with an increase of $\xi$, suggesting localization length increases drastically with an onset of magnetism at low temperature.

\subsection{Magnetoresistance}
To further understand electron transport behavior, we have measured isothermal sheet resistance as a function of magnetic field (up to 85 kOe) at 2 K where the field has been applied parallel to the plane of film ($H || ab$-plane). The resistance ($R$) has been measured with field applied in both positive and negative directions. The percentage magnetoresistance (MR), $\Delta R / R(0)$ = $[R(H)-R(0)]/R(0)$ $\times$ 100, has been calculated from measured resistance for present SrIrO$_3$ film at 2 K and shown in Fig. 6 with positive field direction. The film shows positive MR i.e., resistance increases with applied magnetic field. The calculated value of MR is $\sim$ 0.33 \% at 80 kOe, which is not significant but agrees with other reports.\cite{biswas2} We, however, do not observe any cusp/peak in MR close to zero field which is usually seen due to weak antilocalization effect in strong SOC systems. Interestingly, a hysteresis in MR data between increasing and decreasing field has been observed below around 40 kOe field. This hysteresis in MR appears to be connected with the magnetic state of film at low temperature, as similarly a remnant magnetization has been observed in $M(H)$ plot (Fig. 4). With application of high magnetic field, there is an induced FM moment which is retained even after applied field returns to zero. Given that FM spin ordering has significant effect on charge transport behavior (Fig. 5), the difference in moment between increasing and decreasing field causes hysteresis in MR. In high field regime ($>$ 40 kOe), hysteresis is not evident due to magnetic saturation. To check further, we have plotted MR as a function of $M^2$ in inset of Fig. 6. The MR follows a linear behavior with square of magnetization in field range up to $\sim$ 5 kOe, after that the increase of moment does not scale with that of MR. This is clear in Fig. 4 which shows a faster increase of moment up to $\sim$ 5 kOe and after that the rate of increase in moment slows down. In low field regime (below $\sim$ 40 kOe), the MR follows a quadratic field dependance, MR $\propto$ $B^2$, as shown in Fig. 6 with solid line.

\begin{figure}
\centering
		\includegraphics[width=8cm]{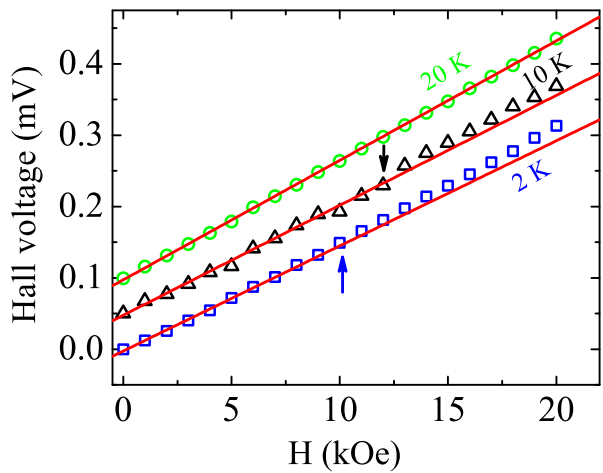}
\caption{(color online) Hall voltage vs magnetic field are shown at three different temperatures i.e., 2, 10 and 20 K. The data at 10 and 20 K are vertically shifted by 0.1 and 0.2 mV, respectively for clarity. The vertical arrows indicate the deviation from linearity at respective fields.}
	\label{fig:Fig7}
\end{figure}

\subsection{Hall measurements}
In an aim to investigate the nature of charge carriers and magnetic state, Hall voltage has been measured as a function of magnetic field in low temperature magnetic state at 2, 10 and 20 K with $H || ab$-plane. Fig. 7 shows Hall voltage is linear with field (up to 20 kOe) at 20 K but with lowering in temperature a nonlinearity is introduced where the onset field for nonlinearity decreases with decreasing temperature. Here, it can be noted that FM state in present film dominates or $M(T)$ shows steep rise below around 20 K (Fig. 3). A nonlinear Hall effect is an intrinsic phenomenon which can be explained with various theoretical descriptions where the two-carrier transport model and anomalous Hall effect (AHE) are commonly noted ones. Considering that the film develops a FM ordering at low temperature (Fig. 3) and MR follows a quadratic field dependance signifying a single carrier charge transport (Fig. 6), we have focused on AHE to understand the present Hall effect behavior. In general, Hall resistivity $\rho_{xy}$ can be expressed as,

\begin{eqnarray}
  \rho_{xy} = R_0H + 4\pi R_sM
\end{eqnarray}

\begin{figure}
\centering
		\includegraphics[width=8cm]{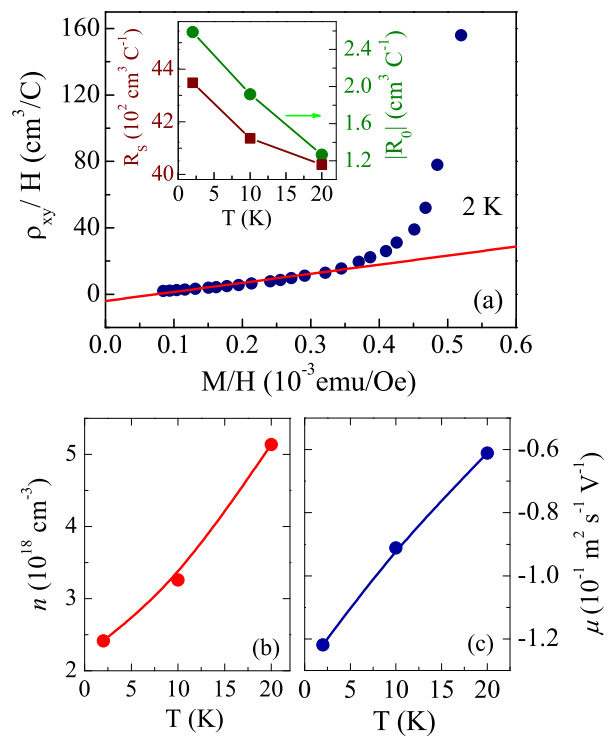}
\caption{(color online) (a) Hall coefficient $R_H$ (= $\rho_{xy}/H$) vs magnetic susceptibility (= $M/H$) are shown at 2 K following Eq. 4 for SrIrO$_3$ film grown on SrTiO$_3$. Inset shows the calculated values of $|R_0|$ and $R_s$ with temperature. (b) and (c) show the variation of carrier concentration $n$ and carrier mobility $\mu$, respectively at low temperatures.}
	\label{fig:Fig8}
\end{figure}

where $H$ is the magnetic field and $M$ is magnetization. While the first term is due to Lorentz force driven ordinary hall resistivity, the second term represents the contribution from anomalous Hall effect. The AHE typically arises in FM materials with broken time-reversal symmetry. While the ordinary Hall effect is due to Lorentz force effect, the origin of anomalous part is debated. Though the mechanisms based on intrinsic or extrinsic (skew-scattering and side jump) scattering are usually thought to cause anomalous Hall effect, the SOC effect plays crucial role in all these mechanisms.\cite{nagaosa} As seen in Eq. 4, AHE is proportional to magnetization, and can effectively be used to investigate the magnetic state of a material. Following Eq. 4, in Fig. 8a we have plotted Hall coefficient $R_H$ (= $\rho_{xy}/H$) as a function of magnetic susceptibility (= $M/H$) at 2 K. A good linear fit to the experimental data is obtained in high field regime, as shown in Fig. 8a. From fitting, we obtain $R_0$ = -2.5881 cm$^3$/C and $R_s$ = 4.35 $\times$ 10$^3$ cm$^3$/C at 2 K. The $R_s$, which is related to magnetic part, is roughly three orders higher than $R_0$ as observed in ordinary magnetic materials. The negative sign of $R_0$ implies an electron like charge carriers in the system. The variation of both $R_0$ and $R_s$ at low temperatures are shown in inset of Fig. 8a, showing both the parameters increases with lowering the temperature. From the relation $R_0$ = -1/$(ne)$, we obtain carrier concentration $n$ = 2.4 $\times$ 10$^{18}$ /cm$^3$ at 2 K, which is in good agreement of earlier reports and indicates low carrier concentration in the system.\cite{zhang1} Mobility of charge carriers has also been evaluated as $\mu$ = $R_0$/$\rho_{xx}$, where $\rho_{xx}$ is the resistivity parallel to the direction of current in presence of zero magnetic field. Calculated values of $n$ and $\mu$ for different temperatures are shown in Fig. 8b and 8c, receptively. Both charge concentration and mobility decreases with decreasing temperature. Nonetheless, the presence of nonlinear Hall effect or anomalous Hall effect confirms a development of FM ordering at low temperature below $\sim$ 20 K in present film. This is in contrast with bulk SrIrO$_3$ which exhibits PM behavior at least down to 1.7 K, though a sharp rise in susceptibility below 15 K indicates this material is in proximity to FM instability.\cite{cao3}

\begin{figure}
\centering
		\includegraphics[width=8cm]{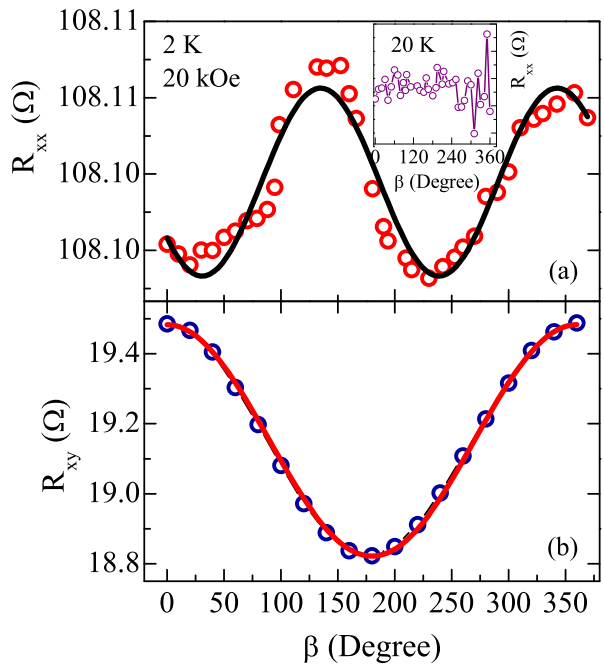}
\caption{(color online) Angular dependance of (a) longitudinal resistance $R_{xx}$ and (b) transverse resistance (planar Hall) are shown for present SrIrO$_3$ film at 2 K with an applied field of 20 kOe. Inset of (a) shows the angular dependence of $R_{xx}$ at 20 K with 20 kOe field.}
	\label{fig:Fig9}
\end{figure}

\subsection{Angle dependent Hall and resistivity measurements}
The low temperature FM state has further been probed by measuring both longitudinal ($R_{xx}$) and transverse ($R_{xy}$) resistance after rotating the film with respect to the direction of applied magnetic field. In case of FM materials with long-range spin ordering, the magnetization has influence on the scattering of carriers and the resistivity depends on angle between magnetization (or magnetic field) and current directions which is usually termed as anisotropic magnetoresistance (AMR) and defined as difference in MR when the current ($I$) is applied either parallel or perpendicular to the magnetization. Figs. 9a and 9b show the measured $R_{xx}$ and $R_{xy}$ as a function of angle $\beta$ between the magnetization and the current direction, respectively. The measurements have been done at 2 K in presence of 20 kOe field where the applied field is in the range of saturation magnetization (see Fig. 4). Though the variation of $R_{xx}$ over angel $\beta$ is not significant but it shows a sinusoidal variation where the $R_{xx}(\beta)$ data are best fitted with the following equation,

\begin{eqnarray}
 R_{xx} = A + B\cos (C\beta + D)
\end{eqnarray}

where $A$ is an offset parameter, $B$ is the amplitude of the angular dependance of MR, $C$ is the multiplying factor to angle and $D$ is the phase factor. Usually, the parameter $C$ takes value 2 but is present case, $C$ = 1.75(3) gives better fitting with phase angle $D$ = 126(5) degree. The variation of $R_{xx}$ with angle $\beta$ implies the film develops FM ordering at low temperature. In inset of Fig. 9a, we have shown the same $R_{xx}(\beta)$ collected at 20 K. As seen in figure, $R_{xx}$ do not exhibit any noticeable angular dependance at 20 K which is in agreement with linear Hall effect evidenced at same temperature (Fig. 7). This is primarily because at 20 K or above the FM ordering is not strong enough to induce an angular dependance or it requires higher magnetic field. We have further recorded planner Hall resistance $R_{xy}$ as a function of angle $\beta$ in hall geometry where current is applied in $xx$ direction and voltage is measured in $xy$ direction. As shown in Fig. 9b, the variation of $R_{xy}$ is also sinusoidal with $\beta$, and can be best fitted using Eq. 4 with $C$ = 1 and negligible phase factor $D$. Nonetheless, the dependance of both $R_{xx}$ and $R_{xy}$ with the angle ($\beta$) between the applied current and magnetic field direction supports a low temperature FM ordering in present film.

\subsection{Temperature dependent structural investigation}
Our previous results indicate an unusual FM state in SrIrO$_3$ film at low temperature which we believe is caused by an enhanced strain or structural distortion. To examine the structural evolution with temperature in present film, temperature dependent XRD measurements have been done down to 20 K. The lattice parameter for both film and substrate have been calculated from Bragg peak. The lattice parameters related to (100) and (200) peaks of SrIrO$_3$ film and (200) peak of SrTiO$_3$ substrate are shown in Figs. 10a, 10b and 10c, respectively.

Here, it can be mentioned that the lattice parameters determined from (100) and (200) peaks of SrIrO$_3$ exhibit a minor difference ($\sim$ 0.003 {\AA}) at room temperature which is may be due to the fact that their 2$\theta$ position is different. Nonetheless, this difference in lattice parameter remains almost similar across the temperature (Fig. 10a and 10b). The Fig. 10 shows that the lattice parameter of film and substrate does not follow the same behavior, particularly at low temperatures. Both the SrIrO$_3$ lattice parameters related to (100) and (200) Bragg peaks, initially decrease with decreasing temperature till around 40 K and then show a sudden increase. On the other hand, lattice parameter of SrTiO$_3$ substrate shows a continuous decrease indicating an anomaly around 100 K (Fig. 10c). This anomaly arises because SrTiO$_3$ has structural phase transition from room temperature cubic to tetragonal phase at 105 K where the lattice parameters show a minor different at low temperature. However, this structural phase transition of SrTiO$_3$ substrate will unlikely influence the magnetic ordering of SrIrO$_3$ film as the former happens at relatively high temperature. As evident in Fig. 10, the difference between lattice parameters of film and substrate increases at low temperature which generates strain resulting in more distortion in local structure. The IrO$_6$ octahedra become distorted in terms of Ir-O bond-length and bond-angle which will have large ramification on the magnetic and electronic properties.

\section{Summary and Conclusion}
The SrIrO$_3$ lies at the boundary of magnetic instability where a close interplay between SOC and $U$ has been shown to give rise many exotic magnetic and electronic phases in this material.\cite{zeb} Even, realization of topological phases through tuning of SOC in SrIrO$_3$ has been discussed using band structure calculations.\cite{carter} The semi-metallic electronic structure in SrIrO$_3$, that is characterized by line nodes and small density of states, makes this material quite interesting compared to layered Sr$_2$IrO$_4$ and Sr$_3$Ir$_2$O$_7$ which are both magnetic and insulating. In this Ruddlesden-Popper series based oxide, the corner shared IrO$_6$ octahedra plays a crucial role in stabilizing its physical properties. An increasing lattice strain realized from underlying substrate would introduce a structural distortion at low temperatures in terms of modification of bond-angle and bond-length between Ir and oxygen. Therefore, the IrO$_6$ octahedra will be distorted with a modified local environment which will influence the 3-dimensional network of Ir-O-Ir chain. Usually, both SOC and $U$ are considered to be intrinsic atomic properties which would largely remain unaltered in present case with no change in transition metal or its electronic configuration. The particular combination of SOC and $U$ is not energetically favorable for the onset of magnetic ordering in bulk SrIrO$_3$, even presence of magnetic Ir$^{4+}$ ($J_{eff}$ = 1/2). However, the IrO$_6$ octahedral (structural) distortion with modified bandwidth and electronic structure would compete with these energies. Therefore, a strong competition among SOC, $U$, structural distortion and bandwidth will eventually weaken the effective strength of SOC and promote the magnetic exchange. These would drive the system into different magnetic and electronic states.

The structural distortion driven magnetism has already been evidenced in doped bulk SrIr$_{1-x}$Sn$_x$O$_3$ where the substitution of nonmagnetic, isovalent Sn$^{4+}$ for Ir$^{4+}$ induces a metal-to-insulator transition and an antiferromagnetic transition ($T_N$ $\geq$ 225 K) which has been explained as combined effect of increased spin-spin exchange interaction, decreased SOC and enhanced IrO$_6$ octahedral distortion.\cite{cui} Further, a dimensionality induced magnetic ordering and its relation with resistivity anomaly has been shown in [(SrIrO$_3$)$_m$,(SrTiO$_3$)$_n$] superlatice where the properties are shown to largely depends on both $m$ and $n$.\cite{matsuno1,hao} Very recently, an evolution of magnetic moment as well as electrical properties with structural distortion has been shown in 3$d$-5$d$ double perovskite (Sr$_{1-x}$Ca$_x$)$_2$FeIrO$_6$ using both experimental data and theoretical calculations.\cite{kharkwal3} A complex orbital ordering with AFM spin state has been shown in other double perovskite Sr$_2$CeIrO$_6$ as a competition between SOC, $U$ and structural distortion.\cite{sudipta} Here, it can be noted that in our previous La$_{0.67}$Sr$_{0.33}$MnO$_3$/SrIrO$_3$ multilayer, we have observed an interface magnetic exchange interaction and related exchange bias effect below 40 K. Although a different magnetic state (weak FM ordering) is observed in present film compared to bulk doped SrIr$_{1-x}$Sn$_x$O$_3$ but an increasing structural distortion at low temperature likely triggers the FM ordering in present SrIrO$_3$ film. Nonetheless, iridates in general have delicate balance among different competing energies such as, electron correlation, SOC, crystal field effect, therefore tuning of any parameter leads to modification of electric and magnetic properties.

\begin{figure}
\centering
		\includegraphics[width=8cm]{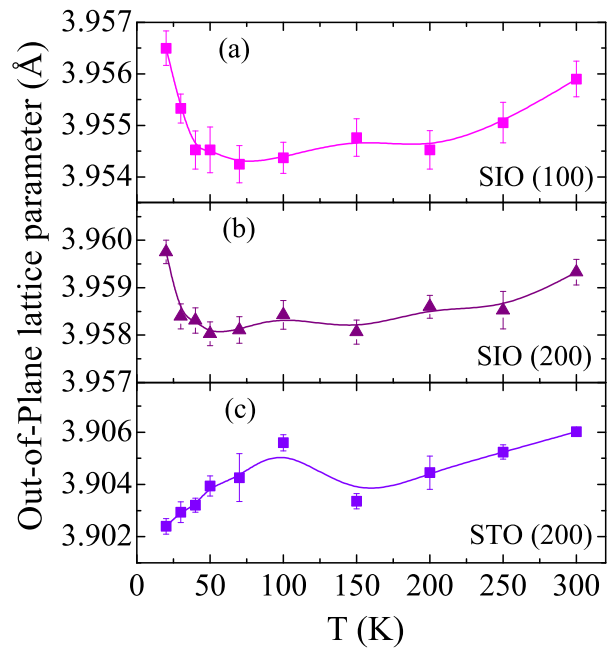}
\caption{(color online) (a) and (b) show the lattice parameter of SrIrO$_3$ film calculated from (100) and (200) Bragg reflection, respectively. (c) shows the lattice parameter of SrTiO$_3$ substrate due to (200) Bragg reflection.}
	\label{fig:Fig10}
\end{figure}

In summary, we have prepared epitaxial thin film of SrIrO$_3$ ($\sim$ 40 nm) on SrTiO$_3$ (100) substrate. Structural analysis shows the film is of good quality. In contrast to bulk material, film shows an insulating behavior where the charge transport mechanisn follows follows Mott's variable-range-hopping model. Further, magnetic measurements suggest a development of weak FM ordering at low temperature below $\sim$ 20 K. The magnetoresistance at 2 K is found to be positive showing a quadratic field dependance in low field regime ($<$ 40 kOe). A nonlinear Hall effect is observed at low temperature below 20 K which is believed to be caused by an anomalous Hall behavior. The present film further shows an anisotropic magnetoresistance and Hall voltage at low temperatures. These experimental observations are in favor of FM state at low temperature. A sudden increase of lattice parameter below $\sim$ 40 K implies an increase of lattice strain which causes (local) structural distortion that is believed to induce the low temperature FM ordering SrIrO$_3$ film. Given that SrIrO$_3$ has delicate energy balance SOC, dimensionality and structural distortion which places this material in close proximity to FM instability and metal-insulator transition, our study shows lattice strain plays a vital role in tuning the physical properties in this simple, though unusual oxide. 

\section{Acknowledgment}
We acknowledge SERB-DST for funding the `Excimer laser', PURSE-DST for funding the `Helium liquefier', FIST-DST for funding the `Low temperature High magnetic field AFM/STM' and UPE II-UGC for funding the `Film deposition chamber'. We are thankful to Dr. Alok Banerjee and Dr. Rajeev Rawat, UGC-DAE CSR, Indore and Dr. Ajay Kr. Shukla, NPL, Delhi for the magnetization, electrical resistivity, thin film XRD measurements and discussions. RC is thankful to UGC, India for the financial support.

\end{document}